\documentclass[journal,9pt]{IEEEtran}
\usepackage{amsmath,amsfonts}
\usepackage{algorithmic}
\usepackage{algorithm}
\usepackage{array}

\usepackage[caption=false,font=normalsize,labelfont=sf,textfont=sf]{subfig}
\usepackage{textcomp}
\usepackage{stfloats}
\usepackage{url}
\usepackage{verbatim}
\usepackage{graphicx}
\usepackage{cite}
\usepackage{hyperref}
\usepackage{amsmath,amssymb,mathtools}
\usepackage{bm}
\usepackage{physics}
\usepackage{tkz-euclide}
\usetikzlibrary{backgrounds}
\usepackage{textcomp}
\usepackage{multirow}
\usepackage{multicol}
\usepackage{refcount}
\usepackage{xfrac}
\usepackage[dvipsnames]{xcolor}
\usepackage{svg}

\usepackage{glossaries}
\setacronymstyle{long-short}

\newacronym{er}{ER}{Effective Roughness}
\newacronym{ds}{DS}{Diffuse Scattering}
\newacronym{mpc}{MPC}{Multipath Components}
\newacronym{pg}{PG}{Path Gain}
\newacronym{sp}{SR}{Specular Reflection}
\newacronym{mut}{MUT}{Material Under Test}
\newacronym{em}{EM}{electromagnetic}
\newacronym{AoA}{AoA}{Angle-of-Arrival}
\newacronym{sage}{SAGE}{Space-Alternating Generalized Expectation-maximization}
\newacronym{xpd}{XPD}{Cross-Polarization Discrimination}
\newacronym{ssa}{SSA}{Singular Spectrum Analysis}
\newacronym{rer}{RER}{Reciprocal ER}
\newacronym{rmsd}{RMSD}{Root Mean Square Deviation}
\newacronym{rmse}{RMSE}{Root Mean Square Error}
\newacronym{go}{GO}{Geometrical Optics}
\newacronym{ka}{KA}{Kirchhoff Approximation}
\newacronym{iem}{IEM}{Integral Equation Method}
\newacronym{spm}{SPM}{Small Perturbation Method}
\newacronym{tx}{TX}{transmitter}
\newacronym{rx}{RX}{receiver}
\newacronym{rcs}{RCS}{Radar Cross Section}
\newacronym{po}{PO}{Physical Optics}
\newacronym{pso}{PSO}{Particle Swarm Optimization}
\newacronym{hh}{HH}{horizontal-horizontal}
\newacronym{hv}{HV}{horizontal-vertical}
\newacronym{vh}{VH}{vertical-horizontal}
\newacronym{vv}{VV}{vertical-vertical}
\newacronym{isac}{ISAC}{Integrated Sensing and Communication}
\newacronym{aoa}{AoA}{Angle of Arrival}
\newacronym{flop}{FLOP}{Floating-point operations}
\newacronym{mse}{MSE}{Mean Square Error}

\loadglsentries{DiffPreamble}

\makeatletter
\def\@IEEEpubidpullup{0.5\baselineskip} % try 2, 2.5, etc.
\makeatother

\def\BibTeX{{\rm B\kern-.05em{\sc i\kern-.025em b}\kern-.08em
    T\kern-.1667em\lower.7ex\hbox{E}\kern-.125emX}}

\begin{document}

\title{A Computationally Efficient Reciprocal Effective Roughness Model for Diffuse Scattering}

\author{Giacomo Melloni, Enrico M. Vitucci,~\IEEEmembership{Senior Member,~IEEE,} Vittorio Degli Esposti,~\IEEEmembership{Senior Member,~IEEE,} Samuel Berweger, Jack Chuang, Camillo Gentile,~\IEEEmembership{Member,~IEEE,}
and Nada Golmie,~\IEEEmembership{Fellow,~IEEE}
%\thanks{Manuscript received April 19, 2021; revised August 16, 2021.}
\thanks{G. Melloni, S. Berweger, J. Chuang, C. Gentile and N. Golmie are with the Communications Technology Laboratory (CTL), National Institute of Standards and Technology (NIST), Gaithersburg, MD 20899 USA (e-mail: \{first.last\}@nist.gov). G. Melloni is also with the Norwegian University of Science and Technology (NTNU), Trondheim, 7034 Norway.}
\thanks{E.M. Vitucci and V. Degli Esposti are with the Department of Electrical, Electronic, and Information Engineering “Guglielmo Marconi” (DEI), CNIT, University of Bologna, 40126 Bologna, Italy (e-mail: \{enricomaria.vitucci,v.degliesposti\}@unibo.it).}
\thanks{(Corresponding author: Giacomo Melloni)}
}

%\markboth{IEEE Transactions on Antennas and Propagation,~Vol.~XX, No. XX, 2025}%
%{Shell %\MakeLowercase{\textit{et al.}}: A Sample Article Using IEEEtran.cls for IEEE Journals}

%\IEEEpubid{0000--0000~\copyright~2025 IEEE}

\maketitle

\begin{abstract}
Ray tracing (RT) has become central to site-specific electromagnetic propagation modeling in complex, dynamic environments. Yet its computational burden grows sharply as high-fidelity digital twins of these environments scale to millions of facets whose material parameters must be continuously updated as the environment changes. The challenge is amplified at mmWave and sub-THz frequencies, where surface roughness becomes comparable to the wavelength and so diffuse scattering can account for up to 40\,\% of the power, making accurate yet tractable models essential. The popular Effective Roughness (ER) approach offers physical consistency but becomes increasingly costly when highly directive lobes are required or when environment parameters change or must be iteratively tuned. This communication introduces a directive, reciprocal diffuse scattering model that preserves the structure of the \gls{er} model while enabling an order-of-magnitude reduction in computational cost. Validation across eight materials shows comparable accuracy, with a slight average improvement, demonstrating a scalable and physically meaningful solution for RT in scenarios where diffuse scattering is relevant.
\end{abstract}


\section{Introduction}
Ray tracing (RT) methods are widely used to model \gls{em} propagation in complex environments, offering efficient predictions based on Geometrical Optics at a fraction of the cost of full-wave solvers. Their adoption has accelerated with GPU-based engines \cite{sionna-rt} and the ability to generate high-fidelity digital replicas of real environments using lidar, camera systems, and AI-driven semantic material labeling \cite{mobicom2026_digital_twin}. These developments have made RT central to emerging digital-twin platforms, where \emph{millions of facets} must be evaluated and—when material parameters are tuned—updated repeatedly \cite{du_realizing_2022}.

As microwave bands saturate, next-generation systems increasingly exploit mmWave and sub-THz frequencies \cite{8373698}. At these bands, surface roughness becomes comparable to the wavelength, making non-specular \gls{ds} a critical propagation mechanism \cite{7491262,9759472} that can contribute up to 40\% of received power~\cite{8319437}. Accurate DS modeling is therefore essential for digital twins operating at 28\,GHz and above \cite{du_realizing_2022}, where industrial and \gls{isac} applications demand reliable predictions for sensing, tracking, and high-capacity communication.

Classical rough-surface scattering has been modeled using physics-based approaches such as the Kirchhoff approximation (KA), the Small Perturbation Method (SPM), and the Integral Equation Method (IEM). While rigorous, these formulations are computationally prohibitive for large scale RT. To overcome this, the \gls{er} model \cite{4052607} has been widely adopted for its efficiency in predicting DS from irregular surfaces, improving RT accuracy across many scenarios \cite{9759472,8288373,8661594,7771383,11084884,Vitucci2019TuningRTmmWave,Vitucci2015UnforgivingValidation}. The recent \gls{rer} model \cite{10137406} strengthens physical consistency by enforcing reciprocity, a fundamental requirement for realistic modeling of radio links \cite{VanBladel2007}.

\IEEEpubidadjcol
A key advantage of the ER family is that \gls{ds} is controlled by only a few parameters such as the normalized scattering coefficient $S \in [0,1]$ that takes into account scattering intensity, and the exponential factor $\alpha_{\scriptscriptstyle R}$ which accounts for the directivity of the scattering lobe, thus enabling efficient calibration to measured data. As a result, the ER model has been widely integrated into commercial RT engines such as \textit{Sionna RT} from NVIDIA\footnote{\label{fn:disclaimer}Certain commercial equipment, instruments, or materials are identified in this paper to foster understanding. Such identification does not imply recommendation or endorsement by the National Institute of Standards and Technology (NIST).} \cite{sionna-rt}, \textit{Wireless InSite} from Remcom\footnotemark[\getrefnumber{fn:disclaimer}], and \textit{Volcano} from Siradel\footnotemark[\getrefnumber{fn:disclaimer}] \cite{SIRADEL_VolcanoFlex_2021}. However, \gls{ds} evaluation becomes computationally demanding in large digital twins, especially when highly directive lobes are required or when material parameters must be iteratively tuned for each facet, as in differentiable RT pipelines \cite{sionna-rt}.

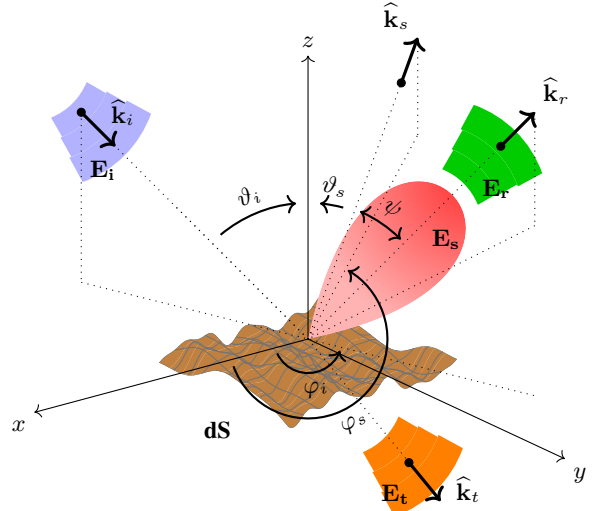
\begin{figure}[!t]
\centering
%====================================================
% Reusable spherical-wavefront object
%====================================================
\pgfkeys{
  /wavefront/.is family, /wavefront,
  color/.store in=\wfcolor,   color=blue!30,
  Rstart/.store in=\wfRstart, Rstart=4.35,
  dR/.store in=\wfdR,         dR=0.20,
  thick/.store in=\wfthick,   thick=0.25,
  overlap/.store in=\wfoverlap, overlap=0.40,
  xscale/.store in=\wfxscale, xscale=1,
  yscale/.store in=\wfyscale, yscale=0.50,
  a1/.store in=\wfa, a1=10,
  a2/.store in=\wfb, a2=11,
  a3/.store in=\wfc, a3=12,
  op1/.store in=\wfopa, op1=0.30,
  op2/.store in=\wfopb, op2=0.50,
  op3/.store in=\wfopc, op3=0.80,
  label/.store in=\wflabel, label=,
  lx/.store in=\wflx, lx=0,
  ly/.store in=\wfly, ly=0
}

\newcommand{\WaveRing}[4]{%
  \path[fill=white, opacity=.9, draw=none]
    ({#1*cos(-#4-\wfoverlap)},{#1*sin(-#4-\wfoverlap)})
    arc[start angle=-#4-\wfoverlap, end angle=#4+\wfoverlap, radius=#1] --
    ({#2*cos(#4+\wfoverlap)},{#2*sin(#4+\wfoverlap)})
    arc[start angle=#4+\wfoverlap, end angle=-#4-\wfoverlap, radius=#2] -- cycle;

  \path[fill=\wfcolor, fill opacity=#3, draw=none]
    ({#1*cos(-#4-\wfoverlap)},{#1*sin(-#4-\wfoverlap)})
    arc[start angle=-#4-\wfoverlap, end angle=#4+\wfoverlap, radius=#1] --
    ({#2*cos(#4+\wfoverlap)},{#2*sin(#4+\wfoverlap)})
    arc[start angle=#4+\wfoverlap, end angle=-#4-\wfoverlap, radius=#2] -- cycle;
}

\newcommand{\WavefrontObject}[2][]{%
  \begin{scope}[#1]
    \pgfkeys{/wavefront,
      color=blue!30,
      Rstart=4.35,
      dR=0.20,
      thick=0.25,
      overlap=0.40,
      xscale=1,
      yscale=0.50,
      a1=10, a2=11, a3=12,
      op1=0.30, op2=0.50, op3=0.80,
      label=,
      lx=0,
      ly=0
    }
    \pgfkeys{/wavefront,#2}

    \begin{scope}[xscale=\wfxscale, yscale=\wfyscale]
      \pgfmathsetmacro{\RinA}{\wfRstart}
      \pgfmathsetmacro{\RoutA}{\wfRstart+\wfthick}
      \WaveRing{\RinA}{\RoutA}{\wfopa}{\wfa}

      \pgfmathsetmacro{\RinB}{\wfRstart+\wfdR}
      \pgfmathsetmacro{\RoutB}{\RinB+\wfthick}
      \WaveRing{\RinB}{\RoutB}{\wfopb}{\wfb}

      \pgfmathsetmacro{\RinC}{\wfRstart+2*\wfdR}
      \pgfmathsetmacro{\RoutC}{\RinC+\wfthick}
      \WaveRing{\RinC}{\RoutC}{\wfopc}{\wfc}

      \node at (\wflx,\wfly) {\wflabel};
    \end{scope}
  \end{scope}
}

\begin{tikzpicture}[scale=1.5]
\path[use as bounding box] (-2.4,-1.3) rectangle (2.6,3.0);
  % --- parameters ---
  \pgfmathsetmacro{\r}{2.5}
  \pgfmathsetmacro{\thetaX}{195} % x-axis direction (deg)
  \pgfmathsetmacro{\thetaY}{335} % y-axis direction (deg)
  \pgfmathsetmacro{\thetaZ}{90}  % z-axis direction (deg)

  % basis vectors (2D projection of a fake-3D frame)
  \pgfmathsetmacro{\cx}{cos(\thetaX)} \pgfmathsetmacro{\sx}{sin(\thetaX)}
  \pgfmathsetmacro{\cy}{cos(\thetaY)} \pgfmathsetmacro{\sy}{sin(\thetaY)}
  \pgfmathsetmacro{\cz}{cos(\thetaZ)} \pgfmathsetmacro{\sz}{sin(\thetaZ)}
  \pgfmathsetmacro{\Ramp}{1.00}% overall roughness amplitude

  % unified surface footprint (use the same for tint + grid)
  \def\Xmin{-2.2} \def\Xmax{2.2}
  \def\Ymin{-1.8} \def\Ymax{1.8}
  \def\bandstep{0.30}   % thickness of gray bands
  \def\gridstep{0.40}   % spacing of grid iso-lines

  % height field
  \pgfmathdeclarefunction{seasurf}{2}{%
    \pgfmathparse{ -0.55
      + \Ramp*(%
          0.18*sin(deg(#1))*cos(deg(#2))
        + 0.12*sin(deg(2.1*#1 + 0.3*#2))
        + 0.15*sin(deg(3.7*#1 - 1.5*#2))
        + 0.10*sin(deg(6.0*#1 + 2.0*#2))
        + 0.06*sin(deg(9.0*#1 - 4.0*#2)))}%
  }

  % --- SURFACE (tint + grid) on background layer ---
  \begin{scope}[on background layer,
    x={(\cx*0.35cm,\sx*0.35cm)},
    y={(\cy*0.35cm,\sy*0.35cm)},
    z={(\cz*0.35cm,\sz*0.35cm)}]

    % --- clip both tint and grid to the same surface boundary ---
    \begin{scope}
      \clip
        % bottom edge: y=Ymin, x from Xmin to Xmax
        plot[samples=120, domain=\Xmin:\Xmax, variable=\xx, smooth]
          ({\xx},{\Ymin},{seasurf(\xx,\Ymin)}) --
        % right edge: x=Xmax, y from Ymin to Ymax
        plot[samples=120, domain=\Ymin:\Ymax, variable=\yy, smooth]
          ({\Xmax},{\yy},{seasurf(\Xmax,\yy)}) --
        % top edge: y=Ymax, x from Xmax to Xmin
        plot[samples=120, domain=\Xmax:\Xmin, variable=\xx, smooth]
          ({\xx},{\Ymax},{seasurf(\xx,\Ymax)}) --
        % left edge: x=Xmin, y from Ymax to Ymin
        plot[samples=120, domain=\Ymax:\Ymin, variable=\yy, smooth]
          ({\Xmin},{\yy},{seasurf(\Xmin,\yy)}) -- cycle;

      % --- soft gray tint (banded fill) ---
      \foreach \yy in {-1.8,-1.5,-1.2,-0.9,-0.6,-0.3,0,0.3,0.6,0.9,1.2,1.5}{
        \pgfmathsetmacro{\yya}{\yy}
        \pgfmathsetmacro{\yyb}{\yy+\bandstep}
        \path[fill=brown, opacity=0.4]
          plot[samples=90, domain=\Xmin:\Xmax, variable=\xx, smooth]
            ({\xx},{\yya},{seasurf(\xx,\yya)})
          --
          plot[samples=90, domain=\Xmax:\Xmin, variable=\xx, smooth]
            ({\xx},{\yyb},{seasurf(\xx,\yyb)})
          -- cycle;
      }

      % --- grid overlay (iso-lines of x and y) ---
      \tikzset{surface grid/.style={line cap=round, line join=round, very thin, black!55, opacity=.65}}

      % constant-y curves
      \foreach \yy in {-1.8,-1.4,-1.0,-0.6,-0.2,0.2,0.6,1.0,1.4,1.8}{
        \draw[surface grid]
          plot[samples=140, domain=\Xmin:\Xmax, variable=\xx, smooth]
            ({\xx},{\yy},{seasurf(\xx,\yy)});
      }

      % constant-x curves
      \foreach \xx in {-2.2,-1.8,-1.4,-1.0,-0.6,-0.2,0.2,0.6,1.0,1.4,1.8,2.2}{
        \draw[surface grid]
          plot[samples=140, domain=\Ymin:\Ymax, variable=\yy, smooth]
            ({\xx},{\yy},{seasurf(\xx,\yy)});
      }

    \end{scope} 

  \end{scope}

  % --- AXES (on top) ---
  \draw[->] (0,0) -- ({\r*cos(\thetaY)},{\r*sin(\thetaY)}) node[below right] {$y$};
  \draw[->] (0,0) -- ({\r*cos(\thetaX)},{\r*sin(\thetaX)}) node[below left]  {$x$};
  \draw[->] (0,0) -- ({\r*cos(\thetaZ)},{\r*sin(\thetaZ)}) node[above]       {$z$};

  % scattered field (red radial shade)
  \shade[shading=radial, inner color=red!80, outer color=red!30, rotate=-45, scale=0.4]
       (0,0) .. controls (-4,6) and (4,6) .. (0,0) -- cycle;

  % Curved arc angles
  \draw[->, thick] (0,0) ++(130:1.2cm) arc (130:95:1.2cm)
    node[midway, sloped, above] {$\vartheta_i$};
  \draw[->, thick] (0,0) ++(75:1.2cm) arc (75:85:1.2cm)
    node[midway, sloped, above] {$\vartheta_s$};
  \draw[->, thick] (0,0) ++({\thetaX+10}:0.3cm) arc ({\thetaX+10}:340:0.3cm)
    node[pos=0.55, sloped, below] {$\varphi_i$};
  \draw[->, thick] (0,0) ++({\thetaX+5}:0.7cm) arc ({\thetaX+5}:420:0.7cm)
    node[pos=0.45, sloped, below] {$\varphi_s$};

\WavefrontObject[
  shift={(-2.70,2.70)},
  rotate=-45
]{color=blue!30, Rstart=0.88, xscale=1, yscale=0.80,
  a1=18, a2=19, a3=20,
  %label={$\mathbf{E_i}$}, 
  lx=-0.9, ly=0.9}

% E_r
\WavefrontObject[
  shift={(0.75,0.75)},
  rotate=45
]{color=green!80!black, Rstart=0.88, xscale=1, yscale=0.80,
  a1=18, a2=19, a3=20,
  label={$\mathbf{E_r}$}, lx=1.05, ly=-0.30}

% E_t
\WavefrontObject[
  shift={(0.05,-0.25)},
  rotate=-45
]{color=orange, Rstart=0.88, xscale=1, yscale=0.80,
  a1=18, a2=19, a3=20,
  label={$\mathbf{E_t}$}, lx=1.30, ly=-0.35}

  % Vectors and labels
  \draw[dotted] (2,2) -- (2,1);
  \draw[dotted] (2,1) -- (0,0);
  \draw[dotted] (-2,2) -- (-2,0.5);
  \draw[dotted] (-2,0.5) -- (2,-0.5);
  \draw[dotted] (0,0) -- (0.9,-1.1);
  \draw[->, very thick] (0.9,-1.1) -- (1.1686827944, -1.4284);
  \filldraw (0.89,-1.09) circle (1pt) node[below, left, xshift=30pt, yshift=-10pt] {$\widehat{\mathbf{k}}_t$};

  \draw[dotted] (0,0) -- (0.9674,1.8);
  \draw[dotted] (0.9674,1.8) -- (0.9674,2.6579);
  
  \draw[dotted] (0,0) -- (-1.18,1.18);

  \node at (-0.8,-0.8) {\textbf{dS}};

  \draw[<->, thick] (0,0) ++(47:1.2cm) arc (45:66:1.2cm)
    node[midway, sloped, above] {$\psi$};

  \draw[dotted] (0,0) -- (0.9674,2.6579) node[above left] {$\widehat{\mathbf{k}}_s$};
  \draw[->, very thick] (0.8223,2.2591) -- (0.9674,2.6579);
  \filldraw (0.8223,2.2591) circle (1pt);

  \draw[dotted] (0,0) -- (2,2) node[above right] {$\widehat{\mathbf{k}}_r$};
  \draw[->, very thick] (1.7,1.7) -- (2,2);
  \filldraw (1.7,1.7) circle (1pt)
    node[above, left, xshift=-12pt, yshift=-35pt] {$\mathbf{E_s}$};

  \draw[dotted] (0,0) -- (-2,2) node[above, right, xshift=0pt, yshift=-22pt] {$\mathbf{E_i}$}; 
  \draw[->, very thick] (-2,2) -- (-1.7,1.7);
  \filldraw (-2,2) circle (1pt) node[above, right,  xshift=8pt, yshift=0pt] {$\widehat{\mathbf{k}}_i$};

\end{tikzpicture}
\caption{Illustration of the \gls{em} fields under consideration, i.e., the incident, reflected, scattered, and transmitted fields ($\mathbf{E_i}$, $\mathbf{E_r}$, $\mathbf{E_s}$, $\mathbf{E_t}$).\label{fig:ReferenceSystem}}
\vspace{-0.5cm}
\end{figure}

In this communication, we propose a new directive model based on the Gaussian scattering function and the reciprocal \gls{er} model in \cite{10137406}. The model resolves three key limitations of existing methods: (1) it achieves high directivity with moderate exponential factors, (2) it allows any positive real exponent, enabling precise lobe tuning, and (3) it is significantly more computationally efficient, yielding an order-of-magnitude speedup over the state-of-the-art with no loss of accuracy. As with the \gls{rer} model, reciprocity is satisfied. Validation across eight materials shows comparable accuracy, with a slight average improvement, demonstrating a scalable and physically meaningful solution for RT in scenarios where diffuse scattering is non-negligible. Due to the lack of space, we will present only one example here.

%%%%%%%%%%%%%%%%%%%%%%%%%%%%%%%%%%%%%%%%%%%%%%%%%%%%
\section{Background: RER Model} \label{sec:back}

\begin{figure*}[!t]
\centering
\subfloat[\label{fig:F_Ratio_Simplification}]{\includegraphics[width=2.2in]{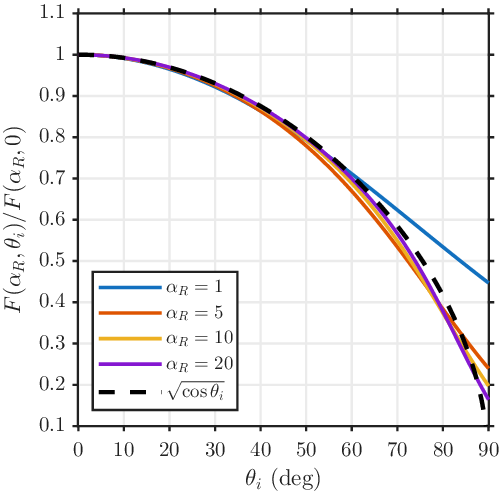}}
\hspace{0.03\textwidth}
\subfloat[\label{fig:Alphas}]{\includegraphics[width=2.1in]
{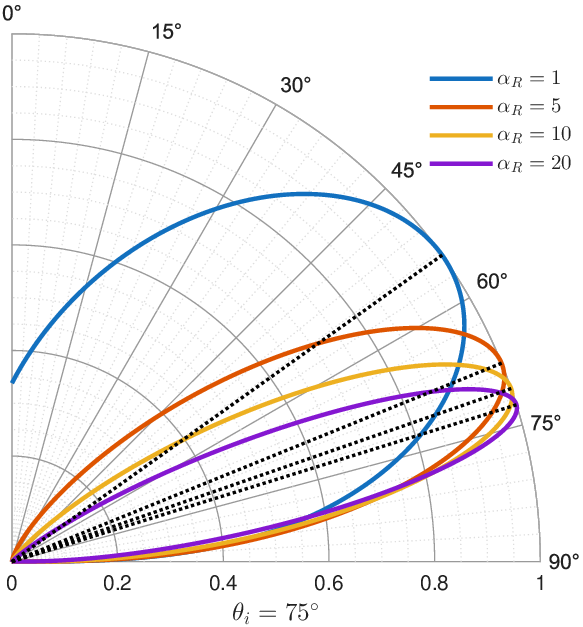}}
\hspace{0.03\textwidth}
\subfloat[\label{fig:Kratio}]{\includegraphics[width=2.2in]
{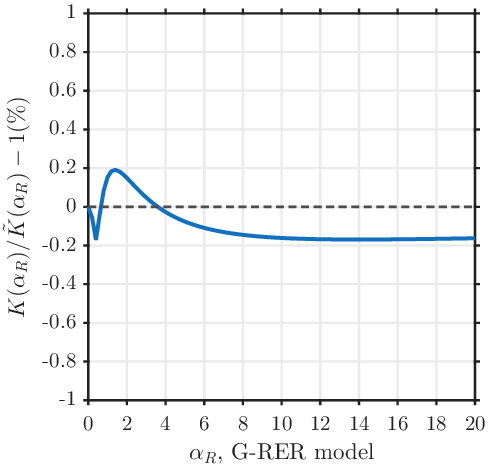}}
\caption{a) Power-balance factor (normalized to its maximum at $\vartheta_i=0^\circ$) \eqref{eq:FalphaRecCloseForm} as a function of $\vartheta_i$, highlighting the $\sqrt{\cos(\vartheta_i)}$ trend; b) Normalized scattering function \eqref{eq:gaussianModelrec} for different exponential values $\alpha_{\scriptscriptstyle R}$ ($\vartheta_i=75^\circ$); c) Ratio between the closed form of $K(\alpha_{\scriptscriptstyle R})$ in \eqref{eq:KalphaCF} and the simplified form in \eqref{eq:Kapprox}.}
\label{fig:Rec_Alphas}
\vspace{-0.5cm}
\end{figure*}

This section provides an overview of the \gls{er} approach and the \gls{rer} model. Its geometrical properties are illustrated in Fig.\,\ref{fig:ReferenceSystem}. First, we define $\hat{\mathbf{k}}_i$, $\hat{\mathbf{k}}_s$, $\hat{\mathbf{k}}_r$ as the incident, scattering, and reflection normalized wavevectors, where the latter provides the direction of the specular reflection, defined as

\begin{equation}
\hat{\mathbf{k}}_r = \hat{\mathbf{k}}_i - 2 \; (\hat{\mathbf{k}}_i \cdot \hat{\mathbf{z}}) \; \hat{\mathbf{z}}
\end{equation}
and $\hat{\mathbf{z}}$ is the normal vector on the surface element dS. As reported in \cite{10137406}, one can show that the \gls{ds} intensity can be written as follows
\begin{equation}
\lvert \mathbf{E}_S \rvert^{2}
= \left( \frac{K\, S}{r_i\, r_s} \right)^{2}
\, \Gamma^{2}\, \cos(\vartheta_i)\,
\,
\frac{f(\hat{\mathbf{k}}_s , \hat{\mathbf{k}}_i)}{F(\hat{\mathbf{k}}_i)}\, \text{dS}
\label{eq:e_s}
\end{equation}
where $r_i$ and $r_s$ are the path lengths from the \gls{tx} to the dS center and the path length from \gls{rx} to dS, respectively. $K$ is a term that depends on the \gls{tx} power and antenna gain \cite{10137406}, while $\Gamma=|\mathbf{E}_r|/|\mathbf{E}_i|$ is the field reflection coefficient, defined as the ratio between the reflected and incident electric field magnitudes and obtained from Fresnel's theory.
In particular, $f(\hat{\mathbf{k}}_s , \hat{\mathbf{k}}_i) = f(\vartheta_i,\varphi_i,\vartheta_s,\varphi_s)$ is the scattering pattern and $F(\hat{\mathbf{k}}_i) = F(\vartheta_i,\varphi_i)$ is the normalization factor that guarantees the power balance, defined as
\begin{equation}\label{eq:F}
    F(\hat{\mathbf{k}}_i)=\int_{0}^{2\pi}\!\int_{0}^{\pi/2}
f(\vartheta_i,\varphi_i,\vartheta_s,\varphi_s)\,\sin(\vartheta_s)\;\dd\vartheta_s\dd\varphi_s
\end{equation}
and the angles $\vartheta_i$, $\varphi_i$, $\vartheta_s$ and $\varphi_s$ are depicted in Fig.\,\ref{fig:ReferenceSystem}.
Here, reciprocity means that the scattering power model is invariant when the incident and observation directions are interchanged. Analytically, this can be expressed as
\begin{equation}\label{eq:rec_condition}
    \frac{f(\hat{\mathbf{k}}_s , \hat{\mathbf{k}}_i)}{F(\hat{\mathbf{k}}_i)} \cdot \cos(\vartheta_i) = \frac{f(\hat{\mathbf{k}}_i , \hat{\mathbf{k}}_s)}{F(\hat{\mathbf{k}}_s)} \cdot \cos(\vartheta_s)
\end{equation}
A solution of $f(\hat{\mathbf{k}}_s , \hat{\mathbf{k}}_i)$ that satisfies reciprocity is derived in \cite{10137406}, which is the \gls{rer} model, and is reported hereafter
\begin{equation}\label{eq:stateoftheart_dm}
    f_{\scriptscriptstyle \text{RER}}(\hat{\mathbf{k}}_s , \hat{\mathbf{k}}_i) = \sqrt{\cos(\vartheta_s)} \cdot \left( \frac{1 + \cos(\psi)}{2} \right)^{\alpha_{\scriptscriptstyle R}}
\end{equation}
where
$\psi$ is the angle between reflection and scattering wavevectors, so that $\cos(\psi) = \hat{\mathbf{k}}_r \cdot  \hat{\mathbf{k}}_s$. Moreover, the power-balance normalization factor $F(\hat{\mathbf{k}}_i)$ in \eqref{eq:e_s} is computed as
\begin{equation}\label{eq:Fdr}
\begin{split}
F_{\scriptscriptstyle \text{RER}}(\hat{\mathbf{k}}_i)
&= \frac{2\pi\,\alpha_{\scriptscriptstyle R}!}{2^{\alpha_{\scriptscriptstyle R}}}
\sum_{j=0}^{\alpha_{\scriptscriptstyle R}} \frac{1}{(\alpha_{\scriptscriptstyle R}-j)!\,(j+1)!!} \\
&\sum_{l=0}^{\left\lfloor j/2 \right\rfloor}
\frac{\cos^{\,j-2l}(\vartheta_i)\,\sin^{\,2l}(\vartheta_i)}{2^{\,l}\,l!\,(j-2l)!!},
\;\;\; 0\le \vartheta_i < \frac{\pi}{2}
\end{split}
\end{equation}

It is observed that the power-balance normalization factor has a $\sqrt{\cos(\vartheta_i)}$ trend, which is a consequence of the reciprocity condition \cite{10137406}. By exploiting this property, a simplified expression for it is derived in \cite{10137406}
\begin{equation}\label{eq:k_alpha_rer}
\begin{cases}
F_{\scriptscriptstyle \text{RER}}(\alpha_{\scriptscriptstyle R},\vartheta_i) \approx K_{\scriptscriptstyle \text{RER}}(\alpha_{\scriptscriptstyle R}) \sqrt{\cos(\vartheta_i)} \\
K_{\scriptscriptstyle \text{RER}}(\alpha_{\scriptscriptstyle R}) = \frac{4\pi}{2^{\alpha_{\scriptscriptstyle R}}}\sum_{j = 0}^{\alpha_{\scriptscriptstyle R}} \binom{\alpha_{\scriptscriptstyle R}}{j} \frac{1}{2 j + 3}
\end{cases}
\end{equation}
where $\binom{\cdot}{\cdot}$ is the binomial operator.
\begin{figure*}
\subfloat[\label{fig:FLOP}]{\includegraphics[width=2.1in]{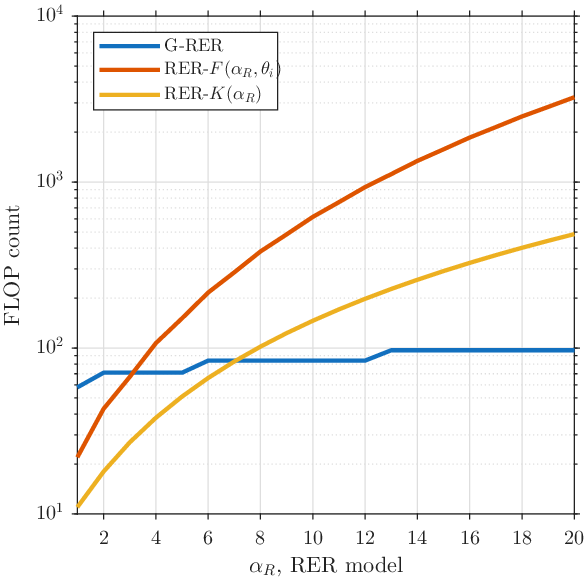}}
\hspace{0.02\textwidth}
\subfloat[\label{fig:AlphasGaussRER}]{\includegraphics[width=2.1in]{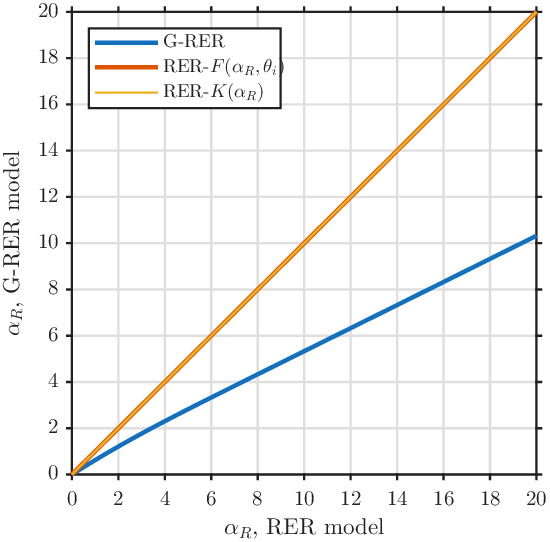}}\hspace{0.02\textwidth}
\subfloat[\label{fig:Patterns}]{\includegraphics[width=2.0in]{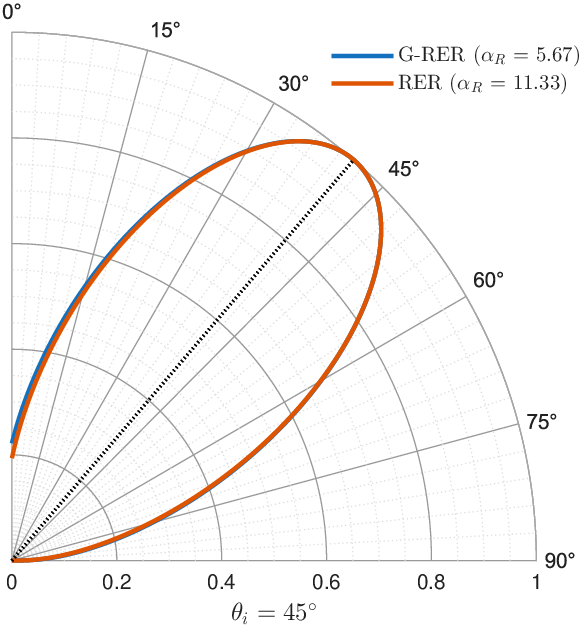}}
\caption{a) Floating-point operations (FLOP) count per facet for different exponential values $\alpha_{\scriptscriptstyle R}$ (blue: using \eqref{eq:KalphaCF}; yellow: using \eqref{eq:k_alpha_rer}; orange: using \eqref{eq:Fdr}); b) Exponential coefficient trend for Gaussian and \gls{rer} models; c) Patterns in \eqref{eq:gaussianModelrec} and \eqref{eq:stateoftheart_dm} for $\varphi_i = \varphi_s = 90^\circ$ used to best fit the measurements.}
\label{fig:metrics}
\vspace{-0.5cm}
\end{figure*}

%%%%%%%%%%%%%%%%%%%%%%%%%%%%%%%%%%%%%%%%%%%%%%%%%%%%
\section{The Proposed Gaussian RER (G-RER) Model}

As previously mentioned, we propose a new model -- which we shall refer to as G-\gls{rer}, where the G stands for Gaussian -- based on a Gaussian scattering function, which is shown below
\begin{equation}\label{eq:gaussianModelrec}
\begin{aligned}
    f_{\scriptscriptstyle \text{G-\gls{rer}}}(\hat{\mathbf{k}}_s , \hat{\mathbf{k}}_i) &= \sqrt{\cos(\vartheta_s)} \cdot \exp \{-\sfrac{1}{2} \; \alpha_{\scriptscriptstyle R} || \hat{\mathbf{k}}_r - \hat{\mathbf{k}}_s ||^2\} \\
    &= \sqrt{\cos(\vartheta_s)} \cdot \exp \{- \alpha_{\scriptscriptstyle R} [1 - \cos(\psi)]\}
\end{aligned}
\end{equation}
Eq.\;\eqref{eq:gaussianModelrec} provides the scattering pattern to be inserted in \eqref{eq:e_s}, instead of \eqref{eq:stateoftheart_dm} which was used in the legacy \gls{rer} model. Moreover, the power-balance normalization factor $F(\hat{\mathbf{k}}_i)$ in \eqref{eq:F} can be computed in closed form according to the following expression (see Appendix)
\begin{equation}\label{eq:FalphaRecCloseForm}
\begin{split}
&F_{\scriptscriptstyle \text{G-\gls{rer}}}(\alpha_{\scriptscriptstyle R},\vartheta_i)
=
2\pi e^{-\alpha_{\scriptscriptstyle R}}\sum_{\ell=0}^{\infty}(2\ell+1)\,i_\ell(\alpha_{\scriptscriptstyle R})\,P_\ell(\cos(\vartheta_i))\,b_\ell
\end{split}
\end{equation}
where $i_\ell(\cdot)$ is the modified spherical Bessel function of the first kind,
$P_{\ell}(\cdot)$ is the Legendre polynomial of degree $\ell$, and $b_\ell$ is a coefficient defined in Appendix.

Due to the presence of the factor $\sqrt{\cos(\vartheta_s)}$ in \eqref{eq:gaussianModelrec}, the reciprocity condition enforced by \eqref{eq:rec_condition} is satisfied only if the following relation holds \cite{10137406}
\begin{equation}
F_{\scriptscriptstyle \text{G-\gls{rer}}}(\alpha_{\scriptscriptstyle R},\vartheta_i) \propto \sqrt{\cos(\vartheta_i)}
\end{equation}
This trend can be clearly observed in Fig.\,\ref{fig:F_Ratio_Simplification}, which shows the normalization factor in \eqref{eq:FalphaRecCloseForm} as a function of the incidence angle $\vartheta_i$ for different values of the exponential factor $\alpha_{\scriptscriptstyle R}$.
Hence, we can approximate the power-balance normalization factor as
\begin{equation}\label{eq:approx}
    F_{\scriptscriptstyle \text{G-\gls{rer}}}(\alpha_{\scriptscriptstyle R},\vartheta_i)\approx K_{\scriptscriptstyle \text{G-\gls{rer}}}(\alpha_{\scriptscriptstyle R})\sqrt{\cos(\vartheta_i)}
\end{equation}
where \(K_{\scriptscriptstyle \text{G-\gls{rer}}}(\alpha_{\scriptscriptstyle R})\) is derived through the least--squares method %(see Appendix \ref{appendix:E})
\begin{equation}\label{eq:least_square}
\begin{split}
K(\alpha_{\scriptscriptstyle R})&=\frac{\langle F(\alpha_{\scriptscriptstyle R},\mu),\,\sqrt{\mu}\rangle}
                 {\langle \sqrt{\mu},\,\sqrt{\mu}\rangle} \\
\end{split}
\end{equation}
with \(\mu=\cos(\vartheta_i)\in[0,1]\), and it yields
\begin{equation}\label{eq:KalphaCF}
\begin{split}
    K_{\scriptscriptstyle \text{G-\gls{rer}}}(\alpha_{\scriptscriptstyle R})&=4\pi\,e^{-\alpha_{\scriptscriptstyle R}}\sum_{\ell=0}^{\infty}(2\ell+1)\,i_\ell(\alpha_{\scriptscriptstyle R})\,b_\ell^{\,2} \\
    &= 16\pi\,e^{-\alpha_{\scriptscriptstyle R}}\sum_{\ell=0}^{\infty}\, \frac{ (2\ell+1)}{(2\ell-1)^2(2\ell+3)^2} \,i_\ell(\alpha_{\scriptscriptstyle R})
\end{split}
\end{equation}
The expression in \eqref{eq:approx} satisfies reciprocity by construction through the term  $\sqrt{\cos(\vartheta_i)}$, similarly to the \gls{rer} model \cite{10137406}. The residual deviation with respect to \eqref{eq:FalphaRecCloseForm} -- below $1\%$ up to $\vartheta_i = 85^\circ$ -- can be considered negligible with respect to the exact power balance.
Fig.\,\ref{fig:Alphas} shows the pattern of the proposed model for different exponential factors and incident angle $\vartheta_i = 75^\circ$. The maximum direction of each lobe (see the dotted lines) tends to the specular reflection direction only for large values of $\alpha_{\scriptscriptstyle R}$ due to the presence of the $\sqrt{\cos(\vartheta_s)}$ term in \eqref{eq:gaussianModelrec}, similarly to the \gls{rer} model \cite{10137406}.

Finally, we propose an approximate solution for $K(\alpha_{\scriptscriptstyle R})$ that does not require evaluating the infinite sum in \eqref{eq:KalphaCF}
\begin{equation}\label{eq:Kapprox}
\widetilde{K}_{\scriptscriptstyle \text{G-\gls{rer}}}(\alpha_{\scriptscriptstyle R})
= \pi\,
\frac{\tfrac{16}{9} + 0.536\,\alpha_{\scriptscriptstyle R} + 0.399\,\alpha_{\scriptscriptstyle R}^{2}}
{1 + 0.965\,\alpha_{\scriptscriptstyle R}
  + 0.457\,\alpha_{\scriptscriptstyle R}^{2} + 0.200\,\alpha_{\scriptscriptstyle R}^{3}}
\end{equation}

The approximation is obtained using the Pad\'e approximant \cite{Datta2004Ch5} for $m=2,n=3$. The coefficients of the Pad\'e approximant were then fitted to \eqref{eq:KalphaCF} by minimizing the maximum relative error over $\alpha_{\scriptscriptstyle R} \in [0, 100]$ using MATLAB\footnotemark[\getrefnumber{fn:disclaimer}]'s \texttt{fminsearch}. Fig.\,\ref{fig:Kratio} depicts the deviation between the approximated normalization factor in \eqref{eq:Kapprox} and \eqref{eq:KalphaCF}, showing a maximum deviation between the two around 0.19\,\% for small values of $\alpha_{\scriptscriptstyle R}$.

This simplification leaves normalized parameters such as RMS delay spread and angular spread unaffected, as the common scalar power factor cancels out in their ratios.
The model also generalizes to a two-lobe scattering pattern \cite{4052607,10137406}, capturing the strong backscattering of certain materials, and can be applied to the monostatic case by setting $\hat{\mathbf{k}}_i = -\hat{\mathbf{k}}_s$.

%%%%%%%%%%%%%%%%%%%%%%%%%%%%%%%%%%%%%%%%%%%%%%%%%%%%
\section{Analysis of Computational Efficiency}

In this section, we compare the computational efficiency of the G-\gls{rer} model with the \gls{rer} model (see Fig.\,\ref{fig:metrics}). To match the beamwidths, we fix $\alpha_{\scriptscriptstyle R}$ for the \gls{rer} and select the G-\gls{rer} exponential factor minimizing the \gls{mse} between patterns~\eqref{eq:stateoftheart_dm} and \eqref{eq:gaussianModelrec} evaluated over all valid off-specular angles $\psi$, via MATLAB\footnotemark[\getrefnumber{fn:disclaimer}]'s \texttt{fminsearch}. Fig.\,\ref{fig:FLOP} shows the per-facet \gls{flop} count for \eqref{eq:Fdr}, \eqref{eq:k_alpha_rer}, and \eqref{eq:KalphaCF}, the series in \eqref{eq:KalphaCF} being truncated once the change between successive terms falls below $\varepsilon_{\text{TH}}=10^{-3}$ (validated against $10^{-20}$). \gls{rer}-F and \gls{rer}-K scale as $\mathcal{O}(\alpha_{\scriptscriptstyle R}^3)$ and $\mathcal{O}(\alpha_{\scriptscriptstyle R}^2)$, whereas G-\gls{rer} stays near $100$ \gls{flop}s across $\alpha_{\scriptscriptstyle R} \in [1,20]$, dropping to $\sim\!10$ with the Pad\'e approximant. Moreover, the series reaches machine precision within $47$ terms, and $\varepsilon_{\text{TH}}=10^{-3}$ keeps the relative error below $0.1\%$. Although \eqref{eq:Kapprox} is intended for RT implementation, the series cost in \eqref{eq:KalphaCF} remains the reference normalization validating the Pad\'e approximation and already yields an order-of-magnitude reduction over \gls{rer}, which \eqref{eq:Kapprox} refines at fixed cost.
Fig.\,\ref{fig:AlphasGaussRER} relates $\alpha_{\scriptscriptstyle R}^{\mathrm{RER}}$ and $\alpha_{\scriptscriptstyle R}^{\mathrm{G\text{-}RER}}$ for $(\varphi_i,\vartheta_i)=(90^\circ,45^\circ)$, each pair found by minimizing the \gls{mse} between the \gls{rer} and G-\gls{rer} patterns over the off-specular angle $\psi$. G-\gls{rer} reproduces the \gls{rer} angular pattern with a smaller exponential factor. This lowers the FLOP count per facet by operating further left on the curve in Fig.\,\ref{fig:FLOP} and narrows the search range for the optimal exponent. Together, these effects significantly accelerate convergence to the optimal exponent.

%%%%%%%%%%%%%%%%%%%%%%%%%%%%%%%%%%%%%%%%%%%%%%%%%%%%
\section{Experimental Validation}
\begin{table}[!t]
\caption{Geometry parameters\label{tab:geom_params}}
\centering
\begin{tabular}{|c||l|c|}
\hline
\textbf{Symbol} & \textbf{Description} & \textbf{Value} \\
\hline
$h_{\scriptscriptstyle \text{RX}}$        & Receiver height            & $1.53~\text{m}$ \\
\hline
$h_{\scriptscriptstyle \text{TX}}$        & Transmitter height         & $1.54~\text{m}$ \\
\hline
$h_{\scriptscriptstyle \text{MUT}}$       & \gls{mut} height        & $1.53~\text{m}$ \\
\hline
$L_{\scriptscriptstyle \text{RX}}$        & RX--\gls{mut} distance           & $1.5~\text{m}$ \\
\hline
$L_{\scriptscriptstyle \text{TX}}$        & TX--\gls{mut} distance           & $0.75~\text{m}$ \\
\hline
$L_{\scriptscriptstyle \text{MUT}}$        & \gls{mut} side length           & $0.6~\text{m}$ \\
\hline
$d$ & Distance offset          & $0.07~\text{m}$ \\
\hline
\end{tabular}
\vspace{-0.3cm}
\end{table}

This section discusses the validation of the Gaussian model in \eqref{eq:gaussianModelrec} with the normalization factor given in \eqref{eq:Kapprox}, using the measurements provided by NIST. We compare the proposed model with the \gls{rer} in \eqref{eq:stateoftheart_dm}. Further details about the measurement setup can be found in \cite{Sloane2025XPD}.

\subsection{Scenario and Measurement Campaign}\label{sec:aaa}
\begin{figure}
\centering
\usetikzlibrary{calc,decorations.pathreplacing}

\newcommand{\ROT}[1][rotate=0]{%
    \tikz [x=0.25cm,y=0.60cm,line width=.3ex,-stealth,#1] \draw (0,0) arc (-150:150:1 and 1);%
}

\newcommand{\TILT}[1][rotate=0]{%
    \tikz [x=0.25cm,y=0.60cm,line width=.3ex,-stealth,#1] \draw (0,0) arc (35:-35:1 and 1);%
}

\begin{tikzpicture}[line cap=round, line join=round,
  dot/.style={circle,fill,inner sep=1.7pt}, scale=0.7]
  \path[use as bounding box] (-3,-0.2) rectangle (10,6.8);

\begin{scope}[xshift=0cm,yshift=0cm]
% all your current drawing commands here

%----------------------- Geometry (units in cm) -----------------------%
% Ground line (just visual reference at y=0)
% \draw[gray!40] (-2.8,0) -- (9.2,0);

% TX mast
\coordinate (TXbase) at (0,-.5);
\coordinate (TX)     at (0,1.0);
\draw (TXbase) -- (TX);
\draw ($(TXbase)+(-.25,0)$) -- ($(TXbase)+(.25,0)$); % little base
\node[below left=0pt and -3pt] at (TXbase) {};

% \node[right] at ($ (TX)!0.55!(TXbase) $) {$h_{\scriptscriptstyle \text{TX}}$};

\draw [decorate,decoration={brace,amplitude=5pt,raise=3pt}]
  ($(TXbase)+(0.0,0.15)$) -- ($(TX)+(0.0,-0.15)$) node[midway, xshift=-20pt,yshift=2pt]{$h_{\scriptscriptstyle \text{TX}}$};

% RX mast
\coordinate (RXbase) at (-2.0,2.5);
\coordinate (RX)     at (-2.0,4.0);
\draw (RXbase) -- (RX);
\draw ($(RXbase)+(-.25,0)$) -- ($(RXbase)+(.25,0)$);

% \node[right] at ($ (RX)!0.48!(RXbase) $) {$h_{\scriptscriptstyle \text{RX}}$};

\draw [decorate,decoration={brace,amplitude=5pt,raise=3pt}]
  ($(RXbase)+(0.0,0.15)$) -- ($(RX)+(0.0,-0.15)$) node[midway, xshift=-20pt,yshift=2pt]{$h_{\scriptscriptstyle \text{RX}}$};
  
% MUT panel (rectangle on the right)
\coordinate (WBL) at (3,1);   % wall bottom-left
\coordinate (WBR) at (7,0.8); % wall bottom-right
\coordinate (WTL) at (3,5);   % wall top-left
\coordinate (WTR) at (7,4.8); % wall top-right
% Trapezoid (your fill)
\draw[thick, fill=brown, opacity=0.8] (WBL) -- (WTL) -- (WTR) -- (WBR) -- cycle;

\begin{scope}
  \clip (WBL) -- (WTL) -- (WTR) -- (WBR) -- cycle;

  \pgfmathtruncatemacro{\N}{6}      % number of subdivisions
  \pgfmathtruncatemacro{\Nm}{\N-1}
  \def\Ext{20} % "infinite" extension factor for clipping
  \foreach \k in {1,...,\Nm}{
    \pgfmathsetmacro{\t}{\k/\N}
    \coordinate (P) at ($(WBL)!\t!(WTL)$); % point on left side
    \draw[very thin, white, opacity=.35]
      ($(P) -\Ext*(WBR) +\Ext*(WBL)$) -- ($(P) +\Ext*(WBR) -\Ext*(WBL)$);
  }
  \foreach \k in {1,...,\Nm}{
    \pgfmathsetmacro{\t}{\k/\N}
    \coordinate (Q) at ($(WBL)!\t!(WBR)$); % point on bottom base
    \draw[very thin, white, opacity=.35]
      ($(Q) -\Ext*(WTL) +\Ext*(WBL)$) -- ($(Q) +\Ext*(WTL) -\Ext*(WBL)$);
  }
\end{scope}

% Point on MUT where rays meet
\coordinate (PMUT) at ($(WTL)!.5!(WBR)$);

% Draw r_i and r_s
\draw[dashed] (TX) -- ($(PMUT)+(-1.65,1.75)$) node[midway,sloped,above] {$r_{\scriptscriptstyle i,n}$};
\draw[dashed] (RX) -- ($(PMUT)+(-1.65,1.75)$) node[midway,sloped,above] {$r_{\scriptscriptstyle s,n}$};

\node[dot,scale=0.5,fill=black, label={[xshift=-30pt,yshift=10pt]right:$\mathrm{dS_n}$}] at ($(PMUT)+(-1.65,1.75)$) {};

% Height/width dimensions (both labelled L_MUT)
\draw[<->] ($(WTL)+(0,.35)$) -- node[above] {$L_{\scriptscriptstyle \text{MUT}}$} ($(WTR)+(0,.35)$);
\draw[<->] ($(WTR)+(.35,0)$) -- node[right] {$L_{\scriptscriptstyle \text{MUT}}$} ($(WBR)+(.35,0)$);

% Rays from TX and RX to PMUT
\draw (TX) -- (PMUT) node[midway,sloped,above] {$L_{\scriptscriptstyle \text{TX}}$};
\draw (RX) -- (PMUT) node[midway,sloped,above] {$L_{\scriptscriptstyle \text{RX}}$};

\node[dot,label={[xshift=-15pt,yshift=-8pt]right:$\mathbf{P}_{\scriptscriptstyle \text{MUT}}$}] at (PMUT) {};

% MUT normal (points left)
\draw[-{Stealth[length=2.5mm]}] (PMUT) -- ++(-1.5,0)
  node[above,pos=.55] {$\hat{\mathbf{n}}_{\scriptscriptstyle \text{MUT}}$};

\draw[-{Stealth[length=2.5mm]}] (PMUT) -- ++(0,0.6)
  node[above,pos=0.8,font=\scriptsize] {$\text{Y}_{\scriptscriptstyle \text{MUT}}$};

\draw[-{Stealth[length=2.5mm]}] (PMUT) -- ++(0.6,-0.03)
  node[above,xshift=2pt,yshift=-1pt,font=\scriptsize] {$\text{X}_{\text{MUT}}$};
  
% Along-surface offset and \mathbf{P}_{T/R}
\coordinate (PTR) at ($(PMUT)+(1.5,0)$);
\draw[solid, draw opacity=0.3] (PMUT) -- (PTR) node[midway, sloped, below, yshift=2pt] {$\scriptscriptstyle d$};

\node[dot,label={[xshift=21pt, yshift=8]left:$\mathbf{P}_{\scriptscriptstyle \text{T / R}}$}] at ($(PTR)+(0,0)$) {};

% MUT mast (for h_MUT near panel)
\coordinate (MUTbase) at ($(PTR)+(0,-2.5)$);
\coordinate (MUTtop)  at ($(PTR)+(0,0)$);
\draw[solid] (MUTbase) -- ($(MUTbase)+(0,0.4)$);
\draw[solid, draw opacity=0.3] (MUTbase) -- (MUTtop);
\draw ($(MUTbase)+(-.25,0)$) -- ($(MUTbase)+(.25,0)$);
% \node[left] at ($ (MUTtop)!0.5!(MUTbase) $) {$h_{\scriptscriptstyle \text{MUT}}$};
\draw [decorate,decoration={brace,amplitude=5pt,raise=3pt}]
  ($(MUTbase)+(0.0,0.15)$) -- ($(MUTtop)+(0.0,-0.15)$) node[midway, xshift=-20pt, yshift=2pt]{$h_{\scriptscriptstyle \text{MUT}}$};

\node[dot,label={[xshift=10pt,yshift=10pt]left:$\mathbf{P}_{\scriptscriptstyle \text{RX}}$},fill=blue] at (RX) {};
\node[dot,label={[xshift=10pt,yshift=10pt]left:$\mathbf{P}_{\scriptscriptstyle \text{TX}}$},fill=blue] at (TX) {};

% Include TILT ROT
\draw[dotted] (PTR) -- ($(PTR)+(0,4)$);
\draw ($(PTR)+(0,3.5)$)  -- ($(PTR)+(0,3.5)$)  node [midway] {\color{green!50!black}\ROT[rotate=-90]} node[midway] {\color{green!50!black}ROT};
\draw[dotted] (PMUT) -- ++(-6.0,0);
\path (PMUT) -- ++(-10.5,0) node[midway] (tilt) {\color{green!50!black}\TILT[rotate=178]} node[midway, xshift=13pt, yshift=-7pt] {\color{green!50!black}TILT};

% \node[dot,label={[xshift=-2pt,yshift=7pt]left:$\mathbf{P}_{\scriptscriptstyle \text{SP}}$}, fill=red] at ($(PMUT)+(0,-1.8)$) {};
\end{scope}
\end{tikzpicture}
\caption{Schematic of the measurement campaign. The distances $L_{\scriptscriptstyle \text{RX}}$ and $L_{\scriptscriptstyle \text{TX}}$ are taken with respect to the center of the antenna systems and the \gls{mut} center when the rotator is at tilt = $0^\circ$ and rotation = $30^\circ$. Given the same configuration, the angle between the \gls{rx} and \gls{tx} pointing directions is $60^\circ$.\label{fig:testbed}}
\vspace{-0.3cm}
\end{figure}
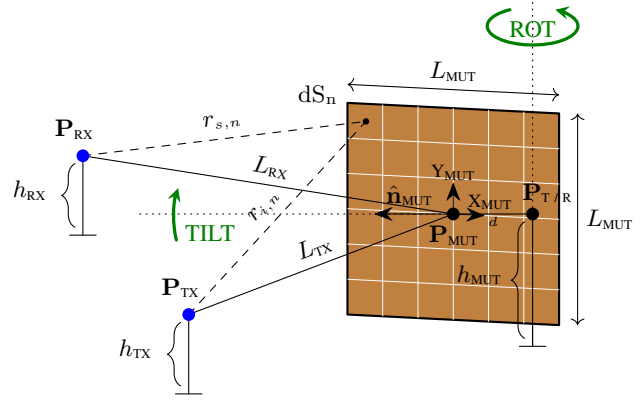

We utilized a fully polarimetric dataset collected at NIST, consisting of bistatic mmWave scattering measurements of eight \gls{mut} samples (see Tab.\,\ref{tab:material_params}), performed at 28.5\,GHz in a semi-anechoic indoor environment with a context-aware channel sounder that combines RF sensing with lidar and camera data for spatial isolation of the \gls{mpc}s scattered by the \gls{mut}. The campaign is designed to characterize the intrinsic scattering properties of individual materials in isolation: all channel contributions from the surrounding environment are removed in post-processing, so the retained response originates solely from the MUT. As shown in Fig.\,\ref{fig:testbed} (geometry in Tab.\,\ref{tab:geom_params}), the \gls{tx} and \gls{rx} are fixed while the \gls{mut} rotates in azimuth ($-15^\circ$ to $75^\circ$) and tilts in elevation ($0^\circ$ to $75^\circ$), yielding 266 orientations, a far wider coverage than common azimuth-only configurations \cite{10137406,8758806}.
Validation here uses tilt $\in [0^\circ, 25^\circ]$ (114 orientations per material), excluding higher tilts where the near-Brewster VV notch in $\Gamma_{\mathrm{VV}}$ is not matched by the measured \gls{ds}.
The sounder transmits a length-2047 pseudo-noise sequence at 2\,Gbps over 2\,GHz bandwidth with 8.8\,dBm power; the receiver samples at 16\,GHz (oversampling factor 8), achieving a 136.4\,dB dynamic range. Both ends employ $8\times8$ dual-polarized phased arrays, the \gls{tx} forming a quasi-omnidirectional beam over 90$^\circ$ azimuth and 50$^\circ$ elevation. Per orientation, 256 channel impulse responses were acquired and \gls{mpc}s estimated via \gls{sage} (0.2$^\circ$ \gls{aoa} and 0.1\,ns delay resolution), with lidar and camera data enabling spatial association. Post-processing removed the LoS path, crosstalk, and environmental scattering, retaining only reflection and diffuse scattering from the \gls{mut} \cite{nist_nextg_repository,9527393,Sloane2025XPD}. For validation, we use the \gls{hh}- and \gls{vv}-channel datasets \cite{nist_nextg_repository}.

\subsection{Methodology}
Each measured \gls{mpc} is classified as either a \gls{ds} or \gls{sp} component. This is possible thanks to the high angular resolution provided by the context-aware channel sounder, allowing us to spatially analyze each \gls{mpc}. To classify the \gls{mpc}s, we used the following algorithm:
\begin{enumerate}
    \item Find the theoretical specular reflection point on the \gls{mut}. Since the target has finite dimensions, the \gls{sp} is observed in only a few orientations;
    \item Identify the \gls{mpc} with the highest path gain intensity $\text{PG}_\text{MAX} = \max\{ \text{PG}_i\}$ inside the circle with radius $\rho$, where $i$ is the $i$-th \gls{mpc} given a certain \gls{mut} orientation;
    \item Assign to the \gls{sp} cluster the $n$-th \gls{mpc} if:
    \begin{itemize}
    \item $\text{PG}_\text{MAX} - \text{PG}_n < \text{PG}_{\text{TH}} $;
    \item the \gls{mpc} is inside the region defined by $\rho$.
    \end{itemize}
\end{enumerate}
Both $\rho = 15$~cm (about twice the projected first Fresnel zone radius, 7.3--8.2~cm) and $\text{PG}_{\text{TH}} = 5$~dB are chosen to avoid overestimating the specular reflection cluster, accounting for the enlarged specular region and reduced specular-to-diffuse intensity separation on rough brick. The complete classification procedure is described in~\cite{melloni2026superresolutionexperimentalvalidationpolarimetric}.
We used NVIDIA's \textit{Sionna RT}\footnotemark[\getrefnumber{fn:disclaimer}] tool \cite{sionna-rt} for calibrating RF parameters, such as the scattering coefficient ($S$), the relative permittivity ($\varepsilon_r$), the conductivity ($\sigma$), and exponential factor $\alpha_{\scriptscriptstyle R}$ for Gaussian and \gls{rer} models. All parameters were estimated using the \gls{pso} method,  with \gls{rmse} loss function using the power in dB values \cite{10137406, 4052607}. The scattering coefficient $S$ was estimated by jointly fitting \gls{sp} and \gls{ds} components, while the optimal $\varepsilon_r$ and $\sigma$ were identified using \gls{sp} total power due to its strength and lower noise compared to \gls{ds}.

To account for the depolarization effect, we used the model in \cite{6183484}, given by
\begin{equation}\label{eq:depol}
\begin{split}
\mathbf{E_{\text{S}}} &= \mathbf{E_{\text{CP,S}}} + \mathbf{E_{\text{XP,S}}} \\
                &= \sqrt{1-\kappa} \cdot |\mathbf{E_{\text{S}}}| \,\hat{\mathbf{n}}_{\scriptscriptstyle \text{CP}}
                       + \sqrt{\kappa}\cdot |\mathbf{E_{\text{S}}}| \,\hat{\mathbf{n}}_{\scriptscriptstyle \text{XP}}\\[6pt]
\end{split}
\end{equation}
where $\mathbf{E_{\text{CP,S}}}$ and $\mathbf{E_{\text{XP,S}}}$ are the total received \gls{ds} fields on the co-polarized and cross-polarized channel, respectively, with the corresponding polarization vectors $\hat{\mathbf{n}}_{\scriptscriptstyle \text{CP}}$ and $\hat{\mathbf{n}}_{\scriptscriptstyle \text{XP}}$, and $\kappa$ is the depolarization coefficient.
We estimated the mean \gls{xpd} using co-polarized (HH, diffuse scattering only) and cross-polarized (HV) channel measurements, obtaining $\text{XPD}_\text{H} \approx$ 17.98\,dB, i.e., $\kappa \approx$\,0.0157.
\begin{figure}[!t]
\centering
\DIFaddbeginFL
\includesvg[width=0.48\textwidth]{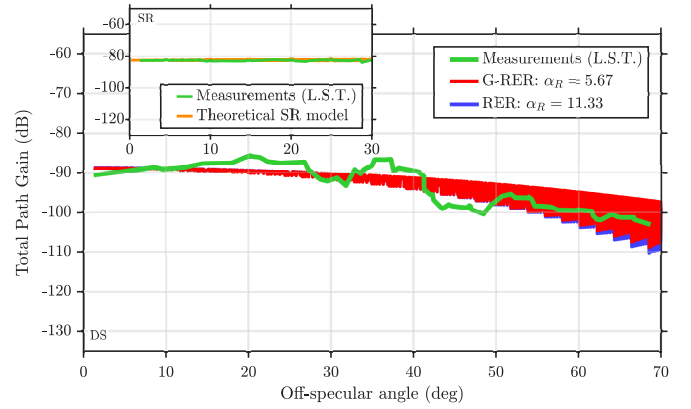}%
\caption{Validation of the G-\gls{rer} in \eqref{eq:gaussianModelrec} and \gls{rer} in \eqref{eq:stateoftheart_dm}, showing the optimal exponential factor $\alpha_{\scriptscriptstyle R}$ in both cases. The green line shows the estimated large scale trend of only the measured \gls{ds} total power.\label{fig:Tot_DS_PG_Brick_HH}}
\vspace{-0.3cm}
\end{figure}
%%%%%%%%%%%%%%%%%%%%%%%%%%%%%%%%%%%%%%%%%%%%%%%%%%%%%%%%%%
\subsection{Results}

This section analyzes the G-\gls{rer} model in \eqref{eq:gaussianModelrec} and the \gls{rer} model in \eqref{eq:stateoftheart_dm} from \cite{10137406}. To estimate $\alpha_{\scriptscriptstyle R}$ more finely for the \gls{rer} model, we used the simplified normalization factor from \cite{10137406}, omitted here for brevity.
Fig.\,\ref{fig:Tot_DS_PG_Brick_HH} compares predicted and measured \gls{ds} total power for the HH channel versus the angle $\psi$ defined in Section\,\ref{sec:back}. The inset shows the corresponding specular reflection power, in very good agreement.
The estimated $\varepsilon_r$ (3.219), $\sigma$ (0.050\,S/m), and $S$ (0.4) agree well with values reported for generic bricks \cite{ITU-R-P2040-3,Fuschini2016ItemLevelMMW,AbelWallace2019_Permittivity_4to40GHz}.
 Power notches occur at maximum tilt, where the \gls{mut} is poorly illuminated, a behavior also predicted by both scattering models. This is visible in the large scale fading (green curve), obtained via an moving-average window of size five, where one corresponds to a single orientation index. The choice follows the relative displacement of \gls{rx} and \gls{tx} with respect to the \gls{mut}, which exceeds $20\lambda$ after five indices, sufficient to filter small-scale fading \cite{1623289}.
Calibration shows a slightly better fit for the G-\gls{rer} compared to the \gls{rer} model: this difference is due to the subtly different shape of the scattering patterns shown in Fig.\,\ref{fig:Patterns}.
Tab.\,\ref{tab:material_params} reports the relative \gls{rmse} results for eight materials and both HH- and VV-polarizations.
A comprehensive polarimetric analysis of the G-\gls{rer} model is provided in \cite{melloni2026superresolutionexperimentalvalidationpolarimetric}.
For the purpose of comparison, we report the relative \gls{rmse} between the G-\gls{rer} and the \gls{rer} model, i.e.,
\[
\mathrm{rRMSE} \; \mathrm{(dB)} =  \mathrm{RMSE} |_{\mathrm{RER}} \; \mathrm{(dB)} - \mathrm{RMSE} |_{\mathrm{G-\gls{rer}}} \; \mathrm{(dB)}
\]
The \gls{rmse} analysis between the two models indicates that the proposed model maintains an accuracy level comparable to that obtained for the \gls{rer}, with a slight improvement observed for the G-\gls{rer}. It should be emphasized that the G-\gls{rer} model is validated using 114 distinct \gls{mut} orientations across eight materials, resulting in 912 independent measurements.
Overall, the Gaussian model offers two main advantages: 1) it requires a smaller $\alpha_{\scriptscriptstyle R}$ than the \gls{rer} model, thus reducing the computational cost; 2) the Pad\'e approximant in \eqref{eq:Kapprox} is efficient, eliminating summations in the normalization factor and further reducing simulation time. This is crucial for ray tracing in complex environments, where diffuse scattering from irregular surfaces is an important propagation mechanism.

%%%%%%%%%%%%%%%%%%%%%%%%%%%%%%%%%%%%%%%%%%%%%%%%%%%%
\section{Conclusion}
This paper proposes an improved directive, reciprocal \gls{ds} model derived from a Gaussian scattering function and validates it using high-resolution measurements for the first time. Compared with the \gls{rer} formulations, the proposed pattern achieves high directivity with smaller exponential factors and admits any positive real exponent, enabling finer calibration. We provided a closed form normalization via spherical Bessel/Legendre expansions and a Pad\'e approximation that removes infinite sums, making the model faster for complex scenarios.
Since ER-type diffuse scattering models are already implemented in several commercial RT tools through an angular scattering pattern, the G-\gls{rer} model can be easily included as a custom pattern by updating the existing ER kernel.
Validation on a high-precision mmWave dataset versus ray-tracing showed that the model reproduces the measured \gls{ds}-only trend across all \gls{mut} orientations and reduces the need for large exponents while improving agreement relative to \gls{rer}.
The G-\gls{rer} model exhibits frequency dependence through $\Gamma^2$ and the fitted parameters $S$ and $\alpha_{\scriptscriptstyle R}$. Although the preliminary validation was carried out at mmWave frequencies, extension to other bands such as sub-THz/THz is therefore feasible, but requires material- and frequency-specific recalibration of $S$ and $\alpha_{\scriptscriptstyle R}$.
Future work will focus on a comprehensive model validation across all available materials and polarizations.

\appendix \label{appendix:A}
\begin{table}[!t]
\caption{\gls{ds} power relative \gls{rmse} improvement.}
\label{tab:material_params}
\centering
\renewcommand{\arraystretch}{1.15}%
\begin{tabular}{|c||l|c|}
\hline
\textbf{Material} & \textbf{Description} & \textbf{r\gls{rmse} (dB), HH/VV}\\
\hline
Brick      & Clay-and-mortar facade, ${\sim}1$\,cm   & +0.22 / -0.02 \\
\hline
Shingles   & Asphalt/bitumen roofing, ${\sim}3$\,mm  & -0.03 / -0.22 \\
\hline
Tile       & Ceramic mosaic tile, ${\sim}8$\,mm      & -0.05 / -0.09 \\
\hline
Carpet     & Synthetic flooring, ${\sim}3$\,mm       & -0.08 / +0.08 \\
\hline
Drywall    & Gypsum panel, ${\sim}12$\,mm            & +0.14 / +0.02 \\
\hline
Mortar     & Tile-adhesion mortar, ${\sim}5$\,cm     & +0.00 / +0.43 \\
\hline
Plexiglass & PMMA sheet, ${\sim}10$\,mm              & +0.00 / +0.22 \\
\hline
Plywood    & Commercial plywood, ${\sim}12$\,mm      & +0.05 / -0.07 \\
\hline
\end{tabular}%
\vspace{-0.3cm}
\end{table}
In what follows, we obtain a closed form expression for the integral in \eqref{eq:F}. This is achieved by expanding the exponential function in \eqref{eq:gaussianModelrec} and using it to derive the closed form solution.
We start from the Legendre expansion in \cite[8.534]{GradshteynRyzhik2014}
\[
e^{i m \rho \cos(\psi)}
=
\sqrt{\frac{\pi}{2 m \rho}}
\sum_{k=0}^{\infty} \mathrm{i}^{\,k} (2k+1)\, J_{k+\tfrac12}(m\rho)\, P_k(\cos(\psi))
\]
where $\mathrm{i}=\sqrt{-1}$, and the following identity holds
\[
J_\nu(\mathrm{i} x)=\mathrm{i}^{\,\nu}\, I_\nu(x)
\]
where $J_\nu( \cdot )$ and $I_\nu( \cdot )$ are the Bessel function of the first kind and modified Bessel function of the first kind, respectively. By setting
$m = -\mathrm{i} \alpha_{\scriptscriptstyle R}$ and $ \rho = 1$, the Legendre expansion simplifies as follows
\begin{equation}\label{eq:firstsol}
e^{\alpha_{\scriptscriptstyle R}\cos(\psi)}=\sum_{\ell=0}^{\infty}(2\ell+1)\,i_\ell(\alpha_{\scriptscriptstyle R})\,P_\ell(\cos(\psi))
\end{equation}
where $i_\ell(\cdot)$ is the modified spherical Bessel function of the first kind, defined by
\begin{equation}\label{eq:m_s_b_f_k}
i_\ell(x)=\sqrt{\frac{\pi}{2x}}\,I_{\ell+1/2}(x)
\end{equation}
Multiplying \eqref{eq:firstsol} by \(e^{-\alpha_{\scriptscriptstyle R}}\) yields \eqref{eq:gaussianModelrec}. Moreover, thanks to the Legendre addition theorem, we write the Legendre expansion integral
\begin{equation}\label{eq:legAddTheo}
    \int_{0}^{2\pi} P_\ell(\cos(\psi))\,\dd\varphi_s=2\pi\,P_\ell(\cos(\vartheta_i))\,P_\ell(\cos(\vartheta_s))
\end{equation}
Then, we insert \eqref{eq:legAddTheo} in \eqref{eq:firstsol}. By including \eqref{eq:firstsol} in the definition of the normalization factor in \eqref{eq:F} we obtain
\[
\begin{split}
F(\alpha_{\scriptscriptstyle R},\vartheta_i)
=&2\pi e^{-\alpha_{\scriptscriptstyle R}}\sum_{\ell=0}^{\infty}(2\ell+1)\,i_\ell(\alpha_{\scriptscriptstyle R})\,P_\ell(\cos(\vartheta_i)) \\
&\cdot \underbrace{\int_{0}^{\pi/2} P_\ell(\cos(\vartheta_s))\,\sqrt{\cos(\vartheta_s)}\,\sin(\vartheta_s)\,\dd\vartheta_s}_{b_\ell}
\end{split}
\]
Let \(u=\cos(\vartheta_s)\in[0,1]\), \(\dd u=-\sin(\vartheta_s)\,\dd\vartheta_s\). Then, we rewrite the coefficient $b_\ell$
\begin{equation}\label{eq:bl}
  b_\ell=\int_{0}^{1} \sqrt{u} P_\ell(u)\,\dd u
\end{equation}
so that
\begin{equation}\label{eq:FalphaRecFinal}
F(\alpha_{\scriptscriptstyle R},\vartheta_i)=
2\pi e^{-\alpha_{\scriptscriptstyle R}}\sum_{\ell=0}^{\infty}(2\ell+1)\,i_\ell(\alpha_{\scriptscriptstyle R})\,P_\ell(\cos(\vartheta_i))\,b_\ell
\end{equation}
What remains is to derive a closed form solution for $b_\ell$.
We start from Gauss’s representation of the Legendre polynomial
\begin{equation}
P_\ell(x)=\,{}_2F_1\!\Big(-\ell,\,\ell+1;\,1;\,\tfrac{1-x}{2}\Big)
\end{equation}
so that
\begin{equation}
P_\ell(u)=\sum_{k=0}^{\ell}\frac{(-\ell)_k\,(\ell+1)_k}{(k!)^2}
\left(\frac{1-u}{2}\right)^k
\end{equation}
where $(n)_k = n (n + 1)\cdots(n + k - 1)$.
Using this form of $P_\ell$ in \eqref{eq:bl} we evaluate
\begin{equation}
\int_0^1 \sqrt{u}\,(1-u)^k\,du
=\text{B}\!\left(\tfrac{3}{2},k+1\right)
=\dfrac{\Gamma(\frac{3}{2})\,k!}{\Gamma(k+\frac{5}{2})}
=\dfrac{2}{3}\,\dfrac{k!}{\left(\frac{5}{2}\right)_k}
\end{equation}
where B$(\cdot,\cdot)$ is the Beta function. Then, $b_\ell$ can be written as follows
\begin{equation}
b_\ell=\frac{2}{3}\sum_{k=0}^{\ell}\frac{(-\ell)_k\,(\ell+1)_k}{\left(\frac{5}{2}\right)_k\,k!}
\Big(\tfrac{1}{2}\Big)^k
=\frac{2}{3}\,{}_2F_1\!\Big(-\ell,\,\ell+1;\,\tfrac{5}{2};\,\tfrac{1}{2}\Big)
\end{equation}

\subsection{Recurrence solution}
Contiguous relations for ${}_2F_1$, applied to the upper parameters
$-\ell$ and $\ell+1$ at $z=\tfrac12$, give
\begin{equation}
(2\ell+7)\,F_{\ell+2}+(2\ell-1)\,F_\ell=0
\end{equation}
which is equivalent to the two-step recurrence
\begin{equation}\label{eq:bn+2}
b_{\ell+2}=-\,\frac{2\ell-1}{2\ell+7}\,b_\ell
\end{equation}
with $b_0=\int_0^1 \sqrt{u}\,du=\tfrac{2}{3}$,
$b_1=\int_0^1 u^{3/2}du=\tfrac{2}{5}$, and $\ell\ge0$.

\subsection{Solution for odd and even indices}
For $\ell=2m+1$, \eqref{eq:bn+2} becomes
\begin{equation}
b_{2m+3}=-\,\frac{4m+1}{4m+9}\,b_{2m+1},
\end{equation}
and iterating gives
\begin{equation}\label{eq:b2m+1}
b_{2m+1}=b_1\prod_{r=0}^{m-1}\!\left(-\frac{4r+1}{4r+9}\right)
=(-1)^m\,\frac{2}{(4m+1)(4m+5)}
\end{equation}
For $\ell=2m$, \eqref{eq:bn+2} gives
\begin{equation}\label{eq:b2m}
b_{2m}=b_0\prod_{r=0}^{m-1}\!\left(-\frac{4r-1}{4r+7}\right)
=(-1)^{m-1}\,\frac{2}{(4m-1)(4m+3)}
\end{equation}
In both cases, the denominator can be written in terms of $\ell$ as
\begin{equation}\label{eq:ident}
\begin{cases}
(4m+1)(4m+5) = (2\ell-1)(2\ell+3), & \ell=2m+1,\\[2mm]
(4m-1)(4m+3) = (2\ell-1)(2\ell+3), & \ell=2m.
\end{cases}
\end{equation}
so that
\begin{equation}\label{eq:blsquare}
|b_\ell|=\frac{2}{(2\ell-1)(2\ell+3)}
\quad\Longrightarrow\quad
b_\ell^2=\frac{4}{(2\ell-1)^2(2\ell+3)^2}
\end{equation}
which completes the derivation of $b_\ell$.

\bibliographystyle{IEEEtran}
\bibliography{main}

\vfill

\end{document}